\documentclass[aps,showpacs,reprint,groupedaddress]{revtex4-1}

\usepackage[dvipdfm]{graphicx}

\newcommand{\C}{$\,^{\circ}\mathrm{C}$}

\begin{document}

\title{Spatial Point Analysis of Quantum Dot Nucleation Sites on InAs Wetting
Layer}

\author{Tomoya Konishi}
\email[]{konishi@anan-nct.ac.jp}
\author{Shiro Tsukamoto}
\affiliation{Anan National College of Technology, Anan, Tokushima
774-0017, Japan}

\date{\today}

\begin{abstract}
We perform spatial point analysis of InAs quantum dot nucleation sites
 and surface reconstruction domain pattern on an InAs wetting
 layer, giving insights for quantum dot nucleation mechanism.
An InAs wetting layer grown to 1.5 monolayers in thickness on a GaAs(001)
 substrate has been observed at 300\C\ by using \textit{in situ} scanning
 tunneling microscopy.
The surface exhibits $(1\times3) / (2\times3)$ and  $(2\times4)$
 reconstruction domains.
 A nearest-neighbor analysis finds that point pattern of quantum dot
 precursors was more 
 similar to that of $(1\times3) / (2\times3)$ domains which are specific
 to Ga-rich region.
This provides the evidence that InAs quantum dot nucleation is induced
 by Ga-rich fluctuation within an InAs wetting layer.
\end{abstract}


\maketitle

Quantum dots (QDs) are potentially used in semiconductor laser devices and
single photon sources of quantum computation and quantum
communication~\cite{arakawa82,li01,fiore07,intallura09}.
Although it has been pointed out that highly dense and uniform QD arrays
are essential for the efficiency of the devices, little is known of the
growth mechanism of QDs to control the nucleation sites on a
Stranski-Krastanow (SK) grown wetting layer (WL).
Some atomic-level theoretical studies on dynamics of surface atoms have
been carried out to understand the growth
mechanisms~\cite{kratzer03,ishii03,fujiwara04,ishii05}.
First principle calculations showed that the migration barrier energy of
In adatom on GaAs(001) surface is higher than that on
1ML-InAs/GaAs(001)~\cite{ishii03,fujiwara04}.
Using kinetic Monte Carlo (kMC) simulations~\cite{ishii05}, Tsukamoto
\textit{et al.} found that some migrating In adatoms were captured on
Ga-rich fluctuation, within an In/Ga mixed layer,
to become a nucleation site~\cite{tsukamoto06}.
To the best of our knowledge, however, there has not been reported any
direct evidence that alloy fluctuation becomes a QD nucleation site.
It is still vital to investigate WL surface in an atomic scale, in
particular the surface reconstruction, preceding QD nucleation.

Since surface reconstruction changes microscopically and dynamically
in the course of WL growth~\cite{belk96,belk97,krzyzewski01}, \textit{in
situ} scanning tunneling microscopy (STM) during  molecular beam epitaxy
(MBE) growth at high temperatures, such as  STMBE~\cite{tsukamoto99}, is
one of powerful tools to observe it.
It is reported that fast Fourier transform analysis of atomic-scale
\textit{in situ} STM images of InAs WL on a GaAs(001) substrate, as well
as reflectance high-energy electron diffraction (RHEED) measurements, has
revealed that the surface reconstruction changes from $c(4\times4)$ to
the mixed structure of $(1\times3)/(2\times3)$ domains and $(2\times4)$
domains prior to QD formation~\cite{tsukamoto06}.

\begin{figure}
 \includegraphics[width=0.9\linewidth,clip]{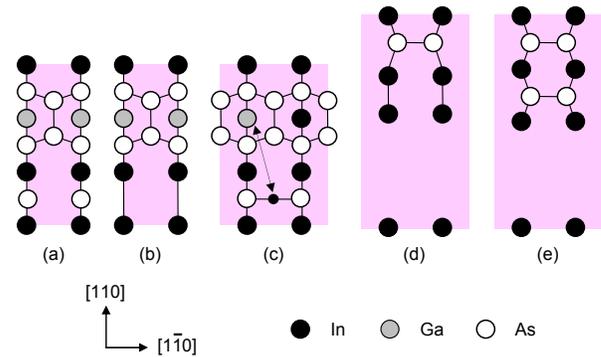}
 \caption{\label{fig:InAs_WL} (Color online) Schematic diagrams of
 surface reconstruction models of InAs/GaAs wetting layer on GaAs(001)
 reproduced from literature: (a) ~$(1\times3)$~\cite{ono01},
 (b) ~$(1\times3)$~\cite{kita02}, (c) ~$(2\times3)$~\cite{ishii03},
 (d) ~$\alpha2(2\times4)$~\cite{yamaguchi95,carter02},
 (e) ~$\beta2(2\times4)$~\cite{yamaguchi95,carter02}.
 Background shade indicates unit cell.
}
\end{figure}

Structure models of $(1\times3)/(2\times3)$ and $(2\times4)$ surface
reconstructions have been investigated by many researchers using
core-level photoemission spectroscopy~\cite{ono01},
reflectance-difference spectroscopy~\cite{kita02}, \textit{ab initio}
calculations in a local density approximation~\cite{kita02,ishii03}, and
STM observations~\cite{yamaguchi95,carter02}, which are, however, still
under discussion.
Figure~\ref{fig:InAs_WL} shows the schematic diagrams, reproduced from
the literature, of some representative surface reconstruction models, in
which only the atoms near the surface are illustrated.
The unit cells of $(1\times3)$ and $(2\times3)$ have one or two of Ga
atoms near the surface whereas those of $(2\times4)$ have none.
In other words, Ga-rich fluctuation in InAs/GaAs WL, which is expected
to become QD nucleation sites, likely forms $(1\times3)/(2\times3)$ surface
reconstruction domains.
It is crucial to investigate the relationship between QD nucleation
sites and surface reconstruction domains, but no researcher has done
yet.
It is quite challenging, however, to witness QD formation specific to
a certain reconstruction domains using typical STM scanning speed
because QD formation occurs on dynamically changing reconstruction,
over a timescale of a few seconds even at very low WL growth rates.
In this paper, we firstly demonstrate a statistical approach to this
problem, namely the statistical comparison of the distribution of surface
reconstruction domains and that of QD nucleation sites.

The distribution of reconstruction domains and QD nucleation sites is
characterized by spatial point patterns; that is a regular (ordered)
pattern, a Poisson (random) pattern, and a clustered (aggregated)
pattern~\cite{diggle03,osullivan03}.
In a regular pattern, points are distributed uniformly.
A Voronoi tessellation, that is polygonal decomposition of a space by
perpendicular bisector lines among neighboring points, is often used
in spatial point analysis.
The standard deviation of Voronoi cell area, $\sigma_\mathrm{Vc}$, represents
point patterns.
Here, we have to be careful of the Voronoi cells touching the boundary of
the study region because of ``edge effect'' that contribution from
points outside cannot be taken into account.
In this study, such \textit{invalid} Voronoi cells were excluded from
 the study region.
For more precise analysis, second-order properties of point patterns
like nearest-neighbor distances are
 useful~\cite{osullivan03,tanemura81,ogata81,baddeley97}.
Let $r$ denote the distance to the nearest point from a randomly selected
location in the study region $R$.
The $F(d)$ function denotes the cumulative frequency distribution of
$r$~\cite{osullivan03} and hence the probability that $r$ occurs less
than any particular distance $d$.
Since the $F$ function is practically identical to the probability of
plotting a random point within any of circles $C_i(d)$, of each radius $d$,
 centered on each of the points, it is simply computed by
\begin{equation}
 F(d)=\frac{\mathrm{Area}\left(\bigcup_{i}
 C_i(d)\cap R \right)}{\mathrm{Area}\left(R\right)},
\end{equation}
where the numerator is the area of the union of the circles, and the
 denominator is the area of the study region $R$.
To compare $F(d)$ between different study regions, $d$ should be
 normalized by the factor $f$ as 
follows:
\begin{equation}
 f = \sqrt{\textrm{Area}\left(R\right)/N},
\end{equation}
where $N$ is the number of points~\cite{tanemura81}.

A piece ($11\times13\times0.6$ mm$^3$) of GaAs(001) crystal was used as
a substrate.
First the surface was thermally cleaned to remove the oxide layer under
$1\times10^{-4}$ Pa of an As$_4$ atmosphere in an MBE growth chamber.
Next, a GaAs buffer layer was grown on the surface by using MBE until
atomically smooth surface was obtained.
The substrate was annealed at 430\C\ for 0.5 h to confirm the formation
of $c(4\times4)$ reconstruction with RHEED.
An STM unit was transferred to the sample holder in the growth chamber
and started scanning.
This means that the sample was neither cooled to room temperature nor
exposed to air to be observed with STM.
The tip bias was -3 V and the tunnel current was 0.2 nA.
A flux of In was irradiated at the InAs growth rate of
$2.5\times10^{-4}$ MLs$^{-1}$.
After 1.5 monolayer (ML) of InAs WL growth, the substrate temperature
was decreased to 300\C\ and the As$_4$ flux was shut off so that the
surface reconstruction should be observed in an atomic scale.
Another samples was prepared in the same way but SK growth of InAs WL
was continued at 430\C\ until QD precursors were formed~\cite{tsukamoto06}.
An STM image of QD precursors was recorded to analyze the distribution of
QD nucleation sites.

\begin{figure}
 \includegraphics[width=0.9\linewidth]{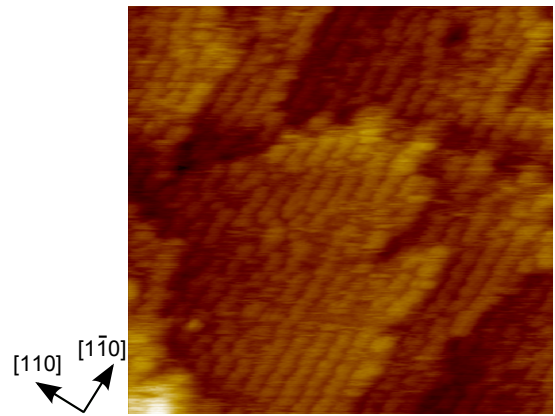}%
 \caption{\label{fig:STM_R4} (Color) 30 nm $\times$ 30 nm filled-state
 STM image of InAs WL on GaAs(001).}
\end{figure}

\begin{figure}
 \includegraphics[width=0.9\linewidth]{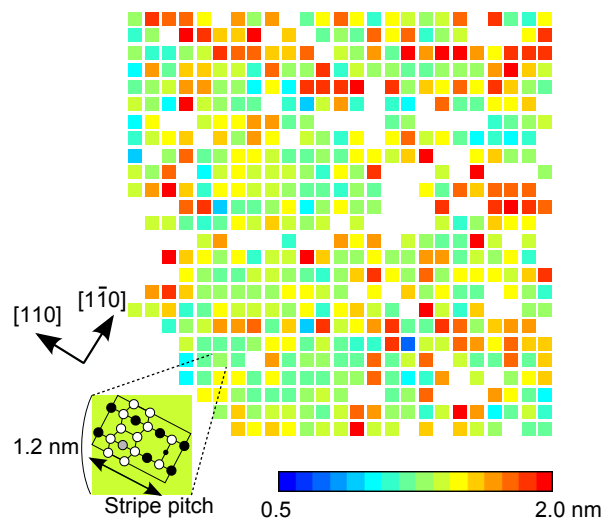}
 \caption{\label{fig:pitch_R4} (Color) Map of As-dimer stripe pitch
 along [110] azimuth measured in each cell of $25\times25$ mesh in
 Fig.~\ref{fig:STM_R4}.
 Schematic diagram of magnified mesh cell is also illustrated aside to
 show the relationship between stripe pitch measured and
 corresponding unit cell of InAs WL surface reconstruction in
 Fig.~\ref{fig:InAs_WL}.
 }
\end{figure}

Figure~\ref{fig:STM_R4} shows the filled-state STM image of 1.5 ML of
InAs WL recorded at 300\C.
The image seems more or less scratched because of migrating In adatoms
on the surface but shows the stripes due to As dimers clearly enough to
measure the pitch.
The stripe pitch corresponds to the unit cell length along [110] azimuth
of InAs WL surface reconstructions in Fig.~\ref{fig:InAs_WL}.
Although it is difficult to discuss the unit cell size of strained and
mixed surface reconstructions, it is expected to have some intermediate
value between those of GaAs and InAs.
We assumed that the stripe pitch along [110] azimuth ranged 0.6--1.0 nm
for $c(4\times4)$, 1.0--1.4 nm for $(1\times3)/(2\times3)$, and
1.4--2.0 nm for $(2\times4)$.
Before measuring the stripe pitch, the STM image [Fig.~\ref{fig:STM_R4}]
was divided by a $25\times25$ mesh.
As shown in the schematic diagram in Fig.~\ref{fig:pitch_R4}, the size
of each mesh cell is 1.2 nm which is comparable to the unit cell sizes
of the InAs WL surface reconstructions.
The stripe pitch was measured from the STM line profile along [110]
azimuth for each mesh cell. 
The data are plotted in the color map of Fig.~\ref{fig:pitch_R4}.
Most cells show $(1\times3)/(2\times3)$ or $(2\times4)$ surface
reconstruction
although some cells are blank because of step edges.

\begin{figure}
 \includegraphics[width=0.9\linewidth]{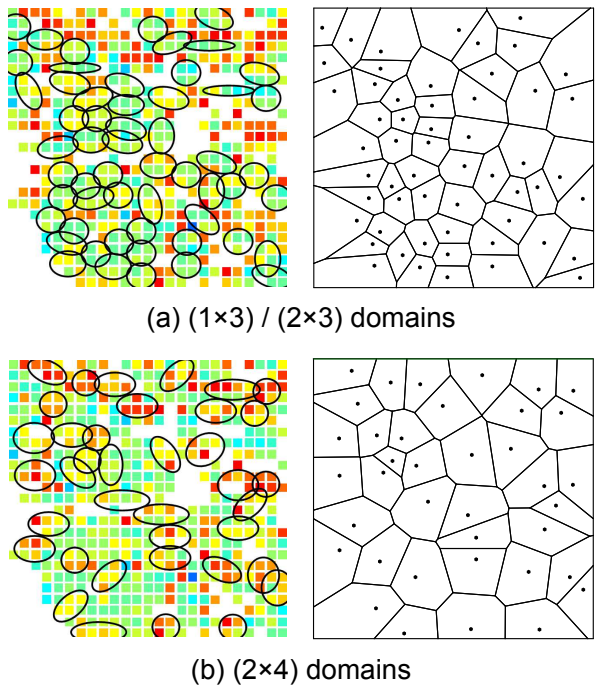}
 \caption{\label{fig:domain_R4} (Color) Surface reconstruction domains,
 indicated by oval markers, and Voronoi tessellations of (a) 
 $(1\times3)/(2\times3)$ and (b) $(2\times4)$ surface reconstruction maps.
 }
\end{figure}

It is predicted that Ga-rich fluctuation, comprised of at least eight Ga
atoms in an In/Ga mixing layer, should be a QD nucleation site according
to kMC simulations~\cite{tsukamoto06}.
For such Ga-rich fluctuation to be formed, four unit cells of
$(1\times3)/(2\times3)$ need to be contiguous, each of which has one or
two Ga atoms.
In this study, groups of four neighboring mesh cells, having the same
surface reconstruction, were located in the map [Fig.~\ref{fig:pitch_R4}]
and indicated by oval markers in Fig.~\ref{fig:domain_R4}.
For each of $(1\times3)/(2\times3)$ and $(2\times4)$ surface
reconstructions, the centroid points of the domains were marked and their
coordinates were measured by using ImageJ
software~\cite{rasband09,abramoff04}.
The centroid coordinates were used for the Voronoi tessellations of the
reconstruction domain maps [Fig.~\ref{fig:domain_R4}] and computation
of the $F$ function~\cite{tanemura81}.

\begin{figure}
 \includegraphics[width=0.9\linewidth]{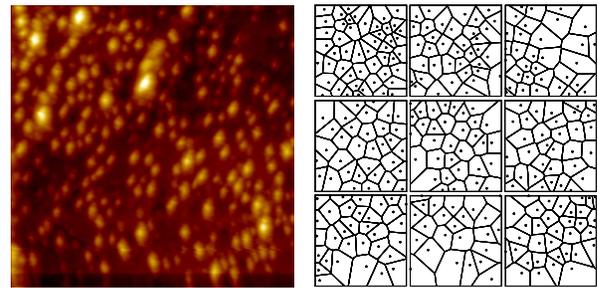}
 \caption{\label{fig:QD} (Color) 150 nm $\times$ 150 nm STM image of QD
 precursors nucleated on SK grown InAs WL at 430\C~\cite{tsukamoto06}
 and Voronoi tessellations computed for each of 3 $\times$ 3 regions.}
\end{figure}

Figure~\ref{fig:QD} shows the 150 nm $\times$ 150 nm STM image of InAs
QD precursors immediately after nucleation~\cite{tsukamoto06}.
The STM image was divided into 3 $\times$ 3 regions.
For each region, measurement of QD coordinates, a Voronoi tessellation,
and computation of the $F$ function were performed as well.

\begin{table}
 \caption{\label{stat_R4} Density $\rho$ and standard deviation of
 Voronoi cell area $\sigma_\mathrm{Vc}$ of surface reconstruction domains
 and QD precursors.}
 \begin{ruledtabular}
  \begin{tabular}{ccc}
   & $\rho$ ($10^{12}$ cm$^{-2}$) & $\sigma_\mathrm{Vc}$\\
   \hline
   $(1\times3)/(2\times3)$ domains & 6.2 & 0.38\\
   $(2\times4)$ domains & 4.2 & 0.31\\
   QD precursors & 0.96--1.7 & 0.20--0.59\\
  \end{tabular}
 \end{ruledtabular}
\end{table}

Table.~\ref{stat_R4} lists the density and standard deviation of
\textit{valid} Voronoi cells computed for the surface reconstruction
domains in Fig.~\ref{fig:domain_R4} and the QD precursors in
Fig.~\ref{fig:QD}.
Each standard deviation is normalized by each study region area.
For QD precursors, the minimum and maximum data in the nine regions are
shown.
The densities of surface reconstruction domains were similar to those of
QD precursors.
The standard deviations of surface reconstruction domains were
in the range of QD precursors.
The similarity in these properties implies some relationship between
surface reconstruction domains and QD precursors. 

\begin{figure}
 \includegraphics[width=\linewidth,clip]{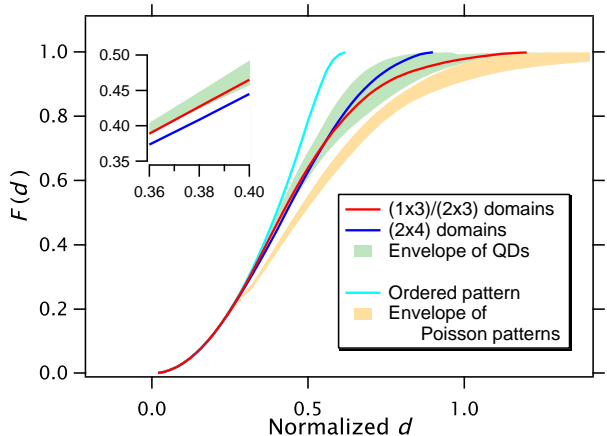}
 \caption{\label{fig:Fd_R4} (Color) Traces of $F$ functions of surface
 reconstruction domains and typical ordered point pattern. Envelope
 regions of QDs in nine regions and typical Poisson
 patterns by accumulating 50 simulations are also shown.
 Magnified view of region where $d$ is 0.36 to 0.40 is superimposed.
}
\end{figure}

The $F$ function will give more precise information.
Figure~\ref{fig:Fd_R4} shows the $F(d)$ traces calculated for
the surface reconstruction domains [~\ref{fig:domain_R4}] and the $F(d)$
envelope of the QD precursors [~\ref{fig:QD}].
The $F(d)$ envelope of typical Poisson patterns was calculated by
accumulating 50 simulations of scattering 50 random points.

Both traces of $(1\times3)/(2\times3)$ and $(2\times4)$ domains were
similar to the envelope of QD precursors although they differs in the
detail.
The traces of the surface reconstruction domains and the QD precursors were
plotted between those of the ordered pattern and the Poisson patterns.
This shows that surface reconstruction domains and QD precursors
are distributed in a rather ordered pattern than a random pattern.
An ordered pattern likely occurs when there is repulsive force among
points.
It is difficult to discuss the origin of repulsive force by STM images,
one can consider the surface strain distribution.
First principles calculations of WL would give some insights for the
surface stain distribution.
It is possible that the uniformity of QD nucleation, which is rather
ordered, is originated from the distribution of surface reconstruction
domains.

The trace of $(2\times4)$ domains is located rather ordered in the QD
envelope, and completely deviates from the QD envelope in small $d$
region as can be seen in the magnified view in Fig.~\ref{fig:Fd_R4}.
The trend of $F(d)$ where $d$ is small is dependent on the pattern in
dense-point areas.
On the other hand, the trend of $F(d)$ where $d$ is large is dependent
on the pattern in sparse-point areas. 
The pattern in dense-point areas is particularly crucial because of
high surface strain which severely affects the distribution.
In the sense of not deviating in dense-point areas,
the trace of $(1\times3)/(2\times3)$ domains represents
the QD precursor envelope better than that of $(2\times4)$ domains.
In other words, the distribution of QD precursors are more similar to
that of $(1\times3)/(2\times3)$ domains than that of $(2\times4)$ domains.
The similarity in the density and the point pattern implies that QD
nucleation is related to $(1\times3)/(2\times3)$ domains where Ga-rich
fluctuation is assumed to be formed.
This also means that QD nucleation sites are already determined at the time
when InAs WL is grown to 1.5 ML or before.

In conclusion, we have shown, by using \textit{in situ} STM, the
similarity in the density and the spatial point patterns between
$(1\times3)/(2\times3)$ surface reconstruction domains on 1.5 ML of
SK grown InAs WL and QD precursors nucleated.
This implies that QD nucleation site is related to the distribution of
$(1\times3)/(2\times3)$ domains at the stage of 1.5 ML-grown InAs WL.
Since a $(1\times3)/(2\times3)$ domain composed of four unit cells is
assumed to contain Ga-rich fluctuation with at least eight Ga atoms,
this model provides a consistent evidence that QD nucleation is induced
by tiny alloy fluctuation as predicted by kMC
simulations~\cite{tsukamoto06}.
The mechanism of QD nucleation, exhibited here, has important
technological implications for the self-assembly and the artificial
arrangement of QDs.

\begin{acknowledgments}
Authors are grateful to Mr. Minoru Yamamoto, Ms. Sayo Yamamoto, and
Mr. Hisanori Iwata.
\end{acknowledgments}


\begin{thebibliography}{10}%
\makeatletter
\providecommand \@ifxundefined [1]{%
 \ifx #1\undefined \expandafter \@firstoftwo
 \else \expandafter \@secondoftwo
\fi
}%
\providecommand \@ifnum [1]{%
 \ifnum #1\expandafter \@firstoftwo
 \else \expandafter \@secondoftwo
\fi
}%
\providecommand \enquote [1]{``#1''}%
\providecommand \bibnamefont  [1]{#1}%
\providecommand \bibfnamefont [1]{#1}%
\providecommand \citenamefont [1]{#1}%
\providecommand\href[0]{\@sanitize\@href}%
\providecommand\@href[1]{\endgroup\@@startlink{#1}\endgroup\@@href}%
\providecommand\@@href[1]{#1\@@endlink}%
\providecommand \@sanitize [0]{\begingroup\catcode`\&12\catcode`\#12\relax}%
\@ifxundefined \pdfoutput {\@firstoftwo}{%
 \@ifnum{\z@=\pdfoutput}{\@firstoftwo}{\@secondoftwo}%
}{%
 \providecommand\@@startlink[1]{\leavevmode}%
 \providecommand\@@endlink[0]{}%
}{%
 \providecommand\@@startlink[1]{%
  \leavevmode
  \pdfstartlink
   attr{/Border[0 0 1 ]/H/I/C[0 1 1]}%
   user{/Subtype/Link/A<</Type/Action/S/URI/URI(#1)>>}%
  \relax
 }%
 \providecommand\@@endlink[0]{\pdfendlink}%
}%
\providecommand \url  [0]{\begingroup\@sanitize \@url }%
\providecommand \@url [1]{\endgroup\@href {#1}{\urlprefix}}%
\providecommand \urlprefix [0]{URL }%
\providecommand \Eprint[0]{\href }%
\@ifxundefined \urlstyle {%
  \providecommand \doi [1]{doi:\discretionary{}{}{}#1}%
}{%
  \providecommand \doi [0]{doi:\discretionary{}{}{}\begingroup
  \urlstyle{rm}\Url }%
}%
\providecommand \doibase [0]{http://dx.doi.org/}%
\providecommand \Doi[1]{\href{\doibase#1}}%
\providecommand \bibAnnote [3]{%
  \BibitemShut{#1}%
  \begin{quotation}\noindent
    \textsc{Key:}\ #2\\\textsc{Annotation:}\ #3%
  \end{quotation}%
}%
\providecommand \bibAnnoteFile [2]{%
  \IfFileExists{#2}{\bibAnnote {#1} {#2} {\input{#2}}}{}%
}%
\providecommand \typeout [0]{\immediate \write \m@ne }%
\providecommand \selectlanguage [0]{\@gobble}%
\providecommand \bibinfo [0]{\@secondoftwo}%
\providecommand \bibfield [0]{\@secondoftwo}%
\providecommand \translation [1]{[#1]}%
\providecommand \BibitemOpen[0]{}%
\providecommand \bibitemStop [0]{}%
\providecommand \bibitemNoStop [0]{.\EOS\space}%
\providecommand \EOS [0]{\spacefactor3000\relax}%
\providecommand \BibitemShut [1]{\csname bibitem#1\endcsname}%
\bibitem{arakawa82}%
  \BibitemOpen
  \bibfield{author}{%
  \bibinfo {author} {\bibfnamefont{Y.}~\bibnamefont{Arakawa}}\ and\ \bibinfo
  {author} {\bibfnamefont{H.}~\bibnamefont{Sakaki}},\ }%
  \bibfield{journal}{%
  \bibinfo {journal} {Appl. Phys. Lett.}\ }%
  \textbf{\bibinfo {volume} {40}},\ \bibinfo {pages} {939} (\bibinfo {year}
  {1982})%
  \bibAnnoteFile{NoStop}{arakawa82}%
\bibitem{li01}%
  \BibitemOpen
  \bibfield{author}{%
  \bibinfo {author} {\bibfnamefont{S.-S.}\ \bibnamefont{Li}}, \bibinfo {author}
  {\bibfnamefont{J.-B.}\ \bibnamefont{Xia}}, \bibinfo {author}
  {\bibfnamefont{J.-L.}\ \bibnamefont{Liu}}, \bibinfo {author}
  {\bibfnamefont{F.-H.}\ \bibnamefont{Yang}}, \bibinfo {author}
  {\bibfnamefont{Z.-C.}\ \bibnamefont{Niu}}, \bibinfo {author}
  {\bibfnamefont{S.-L.}\ \bibnamefont{Feng}},\ and\ \bibinfo {author}
  {\bibfnamefont{H.-Z.}\ \bibnamefont{Zheng}},\ }%
  \bibfield{journal}{%
  \bibinfo {journal} {J. Appl. Phys.}\ }%
  \textbf{\bibinfo {volume} {90}},\ \bibinfo {pages} {6151} (\bibinfo {year}
  {2001})%
  \bibAnnoteFile{NoStop}{li01}%
\bibitem{fiore07}%
  \BibitemOpen
  \bibfield{author}{%
  \bibinfo {author} {\bibfnamefont{A.}~\bibnamefont{Fiore}}, \bibinfo {author}
  {\bibfnamefont{C.}~\bibnamefont{Zinoni}}, \bibinfo {author}
  {\bibfnamefont{B.}~\bibnamefont{Alloing}}, \bibinfo {author}
  {\bibfnamefont{C.}~\bibnamefont{Monat}}, \bibinfo {author}
  {\bibfnamefont{L.}~\bibnamefont{Balet}}, \bibinfo {author}
  {\bibfnamefont{L.~H.}\ \bibnamefont{Li}}, \bibinfo {author}
  {\bibfnamefont{N.~L.}\ \bibnamefont{Thomas}}, \bibinfo {author}
  {\bibfnamefont{R.}~\bibnamefont{Houdr\'e}}, \bibinfo {author}
  {\bibfnamefont{L.}~\bibnamefont{Lunghi}}, \bibinfo {author}
  {\bibfnamefont{M.}~\bibnamefont{Francardi}}, \bibinfo {author}
  {\bibfnamefont{A.}~\bibnamefont{Gerardino}},\ and\ \bibinfo {author}
  {\bibfnamefont{G.}~\bibnamefont{Patriarche}},\ }%
  \bibfield{journal}{%
  \bibinfo {journal} {J. Phys.: Condens. Matter}\ }%
  \textbf{\bibinfo {volume} {19}},\ \bibinfo {pages} {225005} (\bibinfo {year}
  {2007})%
  \bibAnnoteFile{NoStop}{fiore07}%
\bibitem{intallura09}%
  \BibitemOpen
  \bibfield{author}{%
  \bibinfo {author} {\bibfnamefont{P.~M.}\ \bibnamefont{Intallura}}, \bibinfo
  {author} {\bibfnamefont{M.~B.}\ \bibnamefont{Ward}}, \bibinfo {author}
  {\bibfnamefont{O.~Z.}\ \bibnamefont{Karimov}}, \bibinfo {author}
  {\bibfnamefont{Z.~L.}\ \bibnamefont{Yuan}}, \bibinfo {author}
  {\bibfnamefont{P.}~\bibnamefont{See}}, \bibinfo {author}
  {\bibfnamefont{P.}~\bibnamefont{Atkinson}}, \bibinfo {author}
  {\bibfnamefont{D.~A.}\ \bibnamefont{Ritchie}},\ and\ \bibinfo {author}
  {\bibfnamefont{A.~J.}\ \bibnamefont{Shields}},\ }%
  \bibfield{journal}{%
  \bibinfo {journal} {J. Opt. A: Pure Appl. Opt.}\ }%
  \textbf{\bibinfo {volume} {11}},\ \bibinfo {pages} {054005} (\bibinfo {year}
  {2009})%
  \bibAnnoteFile{NoStop}{intallura09}%
\bibitem{kratzer03}%
  \BibitemOpen
  \bibfield{author}{%
  \bibinfo {author} {\bibfnamefont{P.}~\bibnamefont{Kratzer}}, \bibinfo
  {author} {\bibfnamefont{E.}~\bibnamefont{Penev}},\ and\ \bibinfo {author}
  {\bibfnamefont{M.}~\bibnamefont{Scheffler}},\ }%
  \bibfield{journal}{%
  \bibinfo {journal} {Appl. Surf. Sci.}\ }%
  \textbf{\bibinfo {volume} {216}},\ \bibinfo {pages} {436} (\bibinfo {year}
  {2003})%
  \bibAnnoteFile{NoStop}{kratzer03}%
\bibitem{ishii03}%
  \BibitemOpen
  \bibfield{author}{%
  \bibinfo {author} {\bibfnamefont{A.}~\bibnamefont{Ishii}}, \bibinfo {author}
  {\bibfnamefont{K.}~\bibnamefont{Fujiwara}},\ and\ \bibinfo {author}
  {\bibfnamefont{T.}~\bibnamefont{Aisaka}},\ }%
  \bibfield{journal}{%
  \bibinfo {journal} {Appl. Surf. Sci.}\ }%
  \textbf{\bibinfo {volume} {216}},\ \bibinfo {pages} {478} (\bibinfo {year}
  {2003})%
  \bibAnnoteFile{NoStop}{ishii03}%
\bibitem{fujiwara04}%
  \BibitemOpen
  \bibfield{author}{%
  \bibinfo {author} {\bibfnamefont{K.}~\bibnamefont{Fujiwara}}, \bibinfo
  {author} {\bibfnamefont{A.}~\bibnamefont{Ishii}},\ and\ \bibinfo {author}
  {\bibfnamefont{T.}~\bibnamefont{Aisaka}},\ }%
  \bibfield{journal}{%
  \bibinfo {journal} {Thin Solid Films}\ }%
  \textbf{\bibinfo {volume} {464--465}},\ \bibinfo {pages} {35} (\bibinfo
  {year} {2004})%
  \bibAnnoteFile{NoStop}{fujiwara04}%
\bibitem{ishii05}%
  \BibitemOpen
  \bibfield{author}{%
  \bibinfo {author} {\bibfnamefont{A.}~\bibnamefont{Ishii}}, \bibinfo {author}
  {\bibfnamefont{M.}~\bibnamefont{Tsukao}}, \bibinfo {author}
  {\bibfnamefont{N.}~\bibnamefont{Toda}},\ and\ \bibinfo {author}
  {\bibfnamefont{S.}~\bibnamefont{Oshima}},\ }%
  \bibinfo {howpublished} {e-print arXiv:cond-mat/0501233} (\bibinfo {year}
  {2005})%
  \bibAnnoteFile{NoStop}{ishii05}%
\bibitem{tsukamoto06}%
  \BibitemOpen
  \bibfield{author}{%
  \bibinfo {author} {\bibfnamefont{S.}~\bibnamefont{Tsukamoto}}, \bibinfo
  {author} {\bibfnamefont{T.}~\bibnamefont{Honma}}, \bibinfo {author}
  {\bibfnamefont{G.~R.}\ \bibnamefont{Bell}}, \bibinfo {author}
  {\bibfnamefont{A.}~\bibnamefont{Ishii}},\ and\ \bibinfo {author}
  {\bibfnamefont{Y.}~\bibnamefont{Arakawa}},\ }%
  \bibfield{journal}{%
  \bibinfo {journal} {small}\ }%
  \textbf{\bibinfo {volume} {2}},\ \bibinfo {pages} {386} (\bibinfo {year}
  {2006})%
  \bibAnnoteFile{NoStop}{tsukamoto06}%
\bibitem{belk96}%
  \BibitemOpen
  \bibfield{author}{%
  \bibinfo {author} {\bibfnamefont{J.~G.}\ \bibnamefont{Belk}}, \bibinfo
  {author} {\bibfnamefont{J.~L.}\ \bibnamefont{Sudijono}}, \bibinfo {author}
  {\bibfnamefont{D.~M.}\ \bibnamefont{Holmes}}, \bibinfo {author}
  {\bibfnamefont{C.~F.}\ \bibnamefont{McConville}},\ and\ \bibinfo {author}
  {\bibfnamefont{T.~S.}\ \bibnamefont{Jones}},\ }%
  \bibfield{journal}{%
  \bibinfo {journal} {Surf. Sci.}\ }%
  \textbf{\bibinfo {volume} {365}},\ \bibinfo {pages} {735} (\bibinfo {year}
  {1996})%
  \bibAnnoteFile{NoStop}{belk96}%
\bibitem{belk97}%
  \BibitemOpen
  \bibfield{author}{%
  \bibinfo {author} {\bibfnamefont{J.~G.}\ \bibnamefont{Belk}}, \bibinfo
  {author} {\bibfnamefont{C.~F.}\ \bibnamefont{McConville}}, \bibinfo {author}
  {\bibfnamefont{J.~L.}\ \bibnamefont{Sudijono}}, \bibinfo {author}
  {\bibfnamefont{T.~S.}\ \bibnamefont{Jones}},\ and\ \bibinfo {author}
  {\bibfnamefont{B.~A.}\ \bibnamefont{Joyce}},\ }%
  \bibfield{journal}{%
  \bibinfo {journal} {Surf. Sci.}\ }%
  \textbf{\bibinfo {volume} {387}},\ \bibinfo {pages} {213} (\bibinfo {year}
  {1997})%
  \bibAnnoteFile{NoStop}{belk97}%
\bibitem{krzyzewski01}%
  \BibitemOpen
  \bibfield{author}{%
  \bibinfo {author} {\bibfnamefont{T.~J.}\ \bibnamefont{Krzyzewski}}, \bibinfo
  {author} {\bibfnamefont{P.~B.}\ \bibnamefont{Joyce}}, \bibinfo {author}
  {\bibfnamefont{G.~R.}\ \bibnamefont{Bell}},\ and\ \bibinfo {author}
  {\bibfnamefont{T.~S.}\ \bibnamefont{Jones}},\ }%
  \bibfield{journal}{%
  \bibinfo {journal} {Surf. Sci.}\ }%
  \textbf{\bibinfo {volume} {482--485}},\ \bibinfo {pages} {891} (\bibinfo
  {year} {2001})%
  \bibAnnoteFile{NoStop}{krzyzewski01}%
\bibitem{tsukamoto99}%
  \BibitemOpen
  \bibfield{author}{%
  \bibinfo {author} {\bibfnamefont{S.}~\bibnamefont{Tsukamoto}}\ and\ \bibinfo
  {author} {\bibfnamefont{N.}~\bibnamefont{Koguchi}},\ }%
  \bibfield{journal}{%
  \bibinfo {journal} {J. Cryst. Growth}\ }%
  \textbf{\bibinfo {volume} {201--202}},\ \bibinfo {pages} {118} (\bibinfo
  {year} {1999})%
  \bibAnnoteFile{NoStop}{tsukamoto99}%
\bibitem{ono01}%
  \BibitemOpen
  \bibfield{author}{%
  \bibinfo {author} {\bibfnamefont{K.}~\bibnamefont{Ono}}, \bibinfo {author}
  {\bibfnamefont{T.}~\bibnamefont{Mano}}, \bibinfo {author}
  {\bibfnamefont{K.}~\bibnamefont{Nakamura}}, \bibinfo {author}
  {\bibfnamefont{M.}~\bibnamefont{Mizuguchi}}, \bibinfo {author}
  {\bibfnamefont{H.}~\bibnamefont{Kiwata}}, \bibinfo {author}
  {\bibfnamefont{S.}~\bibnamefont{Nakazono}}, \bibinfo {author}
  {\bibfnamefont{K.}~\bibnamefont{Horiba}}, \bibinfo {author}
  {\bibfnamefont{T.}~\bibnamefont{Kihara}}, \bibinfo {author}
  {\bibfnamefont{J.}~\bibnamefont{Okabayashi}}, \bibinfo {author}
  {\bibfnamefont{A.}~\bibnamefont{Kakizaki}},\ and\ \bibinfo {author}
  {\bibfnamefont{M.}~\bibnamefont{Oshima}},\ }%
  \bibfield{journal}{%
  \bibinfo {journal} {Abstracts of ICCG-13/ICVGE-11, Kyoto},\ \bibinfo {pages}
  {402}}%
   (\bibinfo {year} {2001})%
  \bibAnnoteFile{NoStop}{ono01}%
\bibitem{kita02}%
  \BibitemOpen
  \bibfield{author}{%
  \bibinfo {author} {\bibfnamefont{T.}~\bibnamefont{Kita}}, \bibinfo {author}
  {\bibfnamefont{O.}~\bibnamefont{Wada}}, \bibinfo {author}
  {\bibfnamefont{T.}~\bibnamefont{Nakayama}},\ and\ \bibinfo {author}
  {\bibfnamefont{M.}~\bibnamefont{Murayama}},\ }%
  \bibfield{journal}{%
  \bibinfo {journal} {Phys. Rev. B}\ }%
  \textbf{\bibinfo {volume} {66}},\ \bibinfo {pages} {195312} (\bibinfo {year}
  {2002})%
  \bibAnnoteFile{NoStop}{kita02}%
\bibitem{yamaguchi95}%
  \BibitemOpen
  \bibfield{author}{%
  \bibinfo {author} {\bibfnamefont{H.}~\bibnamefont{Yamaguchi}}\ and\ \bibinfo
  {author} {\bibfnamefont{Y.}~\bibnamefont{Horikoshi}},\ }%
  \bibfield{journal}{%
  \bibinfo {journal} {J. Cryst. Growth}\ }%
  \textbf{\bibinfo {volume} {150}},\ \bibinfo {pages} {148} (\bibinfo {year}
  {1995})%
  \bibAnnoteFile{NoStop}{yamaguchi95}%
\bibitem{carter02}%
  \BibitemOpen
  \bibfield{author}{%
  \bibinfo {author} {\bibfnamefont{W.}~\bibnamefont{Barvosa-Carter}}, \bibinfo
  {author} {\bibfnamefont{R.~S.}\ \bibnamefont{Ross}}, \bibinfo {author}
  {\bibfnamefont{C.}~\bibnamefont{Ratsch}}, \bibinfo {author}
  {\bibfnamefont{F.}~\bibnamefont{Grosse}}, \bibinfo {author}
  {\bibfnamefont{J.~H.~G.}\ \bibnamefont{Owen}},\ and\ \bibinfo {author}
  {\bibfnamefont{J.~J.}\ \bibnamefont{Zinck}},\ }%
  \bibfield{journal}{%
  \bibinfo {journal} {Surf. Sci.}\ }%
  \textbf{\bibinfo {volume} {499}},\ \bibinfo {pages} {L129} (\bibinfo {year}
  {2002})%
  \bibAnnoteFile{NoStop}{carter02}%
\bibitem{diggle03}%
  \BibitemOpen
  \bibfield{author}{%
  \bibinfo {author} {\bibfnamefont{P.~J.}\ \bibnamefont{Diggle}},\ }%
  \emph{\bibinfo {title} {Statistical Analysis of Spatial Point Patterns}}\
  (\bibinfo {publisher} {Oxford University Press, Inc.},\ \bibinfo {address}
  {New York},\ \bibinfo {year} {2003})%
  \bibAnnoteFile{NoStop}{diggle03}%
\bibitem{osullivan03}%
  \BibitemOpen
  \bibfield{author}{%
  \bibinfo {author} {\bibfnamefont{D.}~\bibnamefont{O'Sullivan}}\ and\ \bibinfo
  {author} {\bibfnamefont{D.~J.}\ \bibnamefont{Unwin}},\ }%
  \emph{\bibinfo {title} {Geographic Information Analysis}}\ (\bibinfo
  {publisher} {John Wiley \& Sons, Inc.},\ \bibinfo {address} {New Jersey},\
  \bibinfo {year} {2003})%
  \bibAnnoteFile{NoStop}{osullivan03}%
\bibitem{tanemura81}%
  \BibitemOpen
  \bibfield{author}{%
  \bibinfo {author} {\bibfnamefont{M.}~\bibnamefont{Tanemura}}\ and\ \bibinfo
  {author} {\bibfnamefont{Y.}~\bibnamefont{Ogata}},\ }%
  \bibfield{journal}{%
  \bibinfo {journal} {Suuri Kagaku (Mathematical Sciences)}\ }%
  \textbf{\bibinfo {volume} {213}},\ \bibinfo {pages} {11} (\bibinfo {year}
  {1981})%
  \bibAnnoteFile{NoStop}{tanemura81}%
\bibitem{ogata81}%
  \BibitemOpen
  \bibfield{author}{%
  \bibinfo {author} {\bibfnamefont{Y.}~\bibnamefont{Ogata}}\ and\ \bibinfo
  {author} {\bibfnamefont{M.}~\bibnamefont{Tanemura}},\ }%
  \bibfield{journal}{%
  \bibinfo {journal} {Ann. Inst. Statist. Math.}\ }%
  \textbf{\bibinfo {volume} {33}},\ \bibinfo {pages} {315} (\bibinfo {year}
  {1981})%
  \bibAnnoteFile{NoStop}{ogata81}%
\bibitem{baddeley97}%
  \BibitemOpen
  \bibfield{author}{%
  \bibinfo {author} {\bibfnamefont{A.}~\bibnamefont{Baddeley}}\ and\ \bibinfo
  {author} {\bibfnamefont{R.~D.}\ \bibnamefont{Gill}},\ }%
  \bibfield{journal}{%
  \bibinfo {journal} {Ann. Statist.}\ }%
  \textbf{\bibinfo {volume} {25}},\ \bibinfo {pages} {263} (\bibinfo {year}
  {1997})%
  \bibAnnoteFile{NoStop}{baddeley97}%
\bibitem{rasband09}%
  \BibitemOpen
  \bibfield{author}{%
  \bibinfo {author} {\bibfnamefont{W.~J.}\ \bibnamefont{Rasband}},\ }%
  \enquote{\bibinfo {title} {{ImageJ}},}\ \bibinfo {howpublished} {U.\,S.
  National Institutes of Health, Bethesda, Maryland, USA} (\bibinfo {year}
  {1997--2009}),\ \url{http://rsb.info.nih.gov/ij/}%
  \bibAnnoteFile{NoStop}{rasband09}%
\bibitem{abramoff04}%
  \BibitemOpen
  \bibfield{author}{%
  \bibinfo {author} {\bibfnamefont{M.~D.}\ \bibnamefont{Abramoff}}, \bibinfo
  {author} {\bibfnamefont{P.~J.}\ \bibnamefont{Magelhaes}},\ and\ \bibinfo
  {author} {\bibfnamefont{S.~J.}\ \bibnamefont{Ram}},\ }%
  \bibfield{journal}{%
  \bibinfo {journal} {Biophotonics Int.}\ }%
  \textbf{\bibinfo {volume} {11}},\ \bibinfo {pages} {36} (\bibinfo {year}
  {2004})%
  \bibAnnoteFile{NoStop}{abramoff04}%
\end{thebibliography}
\providecommand{\noopsort}[1]{}\providecommand{\singleletter}[1]{#1}%

\end{document}